\documentclass[a4paper,11pt]{article}
\usepackage{pos}

\title{Particle Identification with the ePIC detector at the EIC}

\author*[a]{Chandradoy Chatterjee}
\onbehalf{on behalf of the ePIC collaboration}

\affiliation[a]{INFN sezione di Trieste,\\
  Via Vallerio 2, Trieste, Italy}

\emailAdd{chandradoy.chatterjee@ts.infn.it}

\abstract{The ePIC detector is being designed as a general-purpose detector to deliver the full physics program of the Electron-Ion Collider (EIC) in BNL USA. Particle Identification (PID) plays a crucial role in the EIC physics program. Over a wide phase-space the PID systems provide excellent hadron identification required for the EIC physics program. Additionally, the PID systems have to provide low momentum electron-pion separation to support the calorimeter performance. The ePIC experiment employs different advanced PID technologies to reach the physics goals. The compact set-up the collider and the presence of high magnetic field of the experimental solenoid impose numerous technical challenges. In this article we will describe the PID subsystems of the ePIC detector, with a specific emphasis on high momentum PID system using RICH technology. The article will report about the studies performed with SiPMs as the photo-sensor candidate for forward RICH detector. Also, studies made with novel LAPPD detectors and standard MCP-PMTs chosen as baseline photo-sensor candidates for the backward RICH and barrel DIRC system respectively. This article will include a discussion on the simulation studies made with stand-alone Geant4 simulations and official ePIC software framework.
	
 }

\FullConference{31st International Workshop on Deep Inelastic Scattering (DIS2024)\\
 8–12 April 2024\\
Grenoble, France\\}

\begin{document}
\maketitle

\section{Introduction}

\begin{figure}[!thb]
	\centering
	\includegraphics[width=0.75\textwidth]{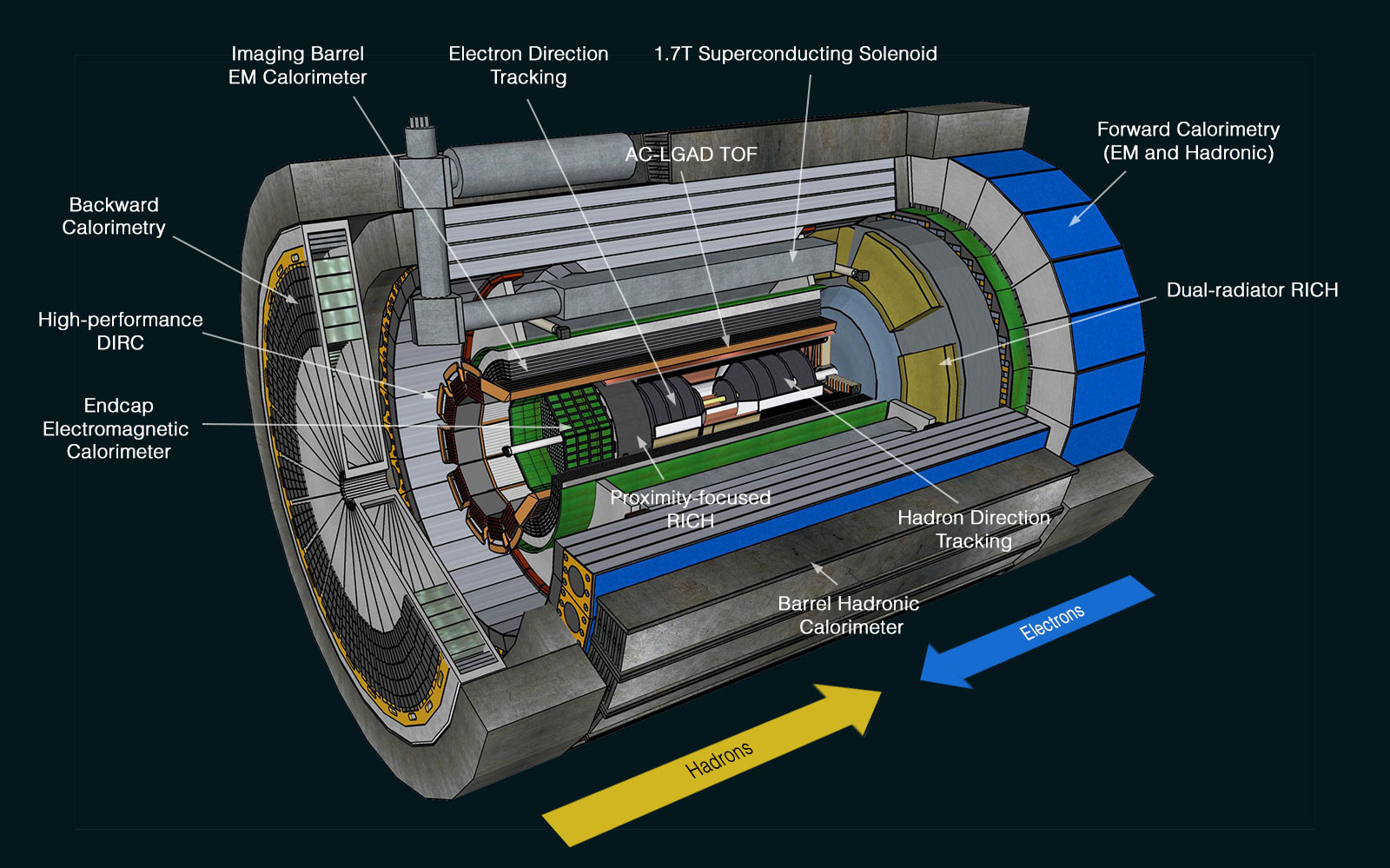}
	\caption{A 3D model of the ePIC detector and PID devices are labeled accordingly.}
	\label{fig:epic_scheme}
\end{figure}

The success of the EIC physics program at the ePIC experiment greatly depends on efficient particle identification. The PID detectors have to play two roles: first, to identify final state charged hadrons and second, to reject pions for reconstruction of event with identified scattered electrons. Given the challenging physics requirements as prescribed by the EIC Yellow report \cite{eicyr} and limited available space, different PID technologies have been identified for the ePIC detector. In the hadron going direction, where most of the high energy hadrons are produced, a dual Radiator Ring Imaging CHerenkov (dRICH) detector will be used to identify electrons, pions, kaons and protons between few hundred MeV/c and 50 GeV/c. It will also provide electron and pion separation from few hundred MeV/c to 10 GeV/c. Apart from the dRICH, in the forward direction a layer of time of flight (TOF) detector is placed to perform electron, pion,kaon and proton identification up to several GeV/c using AC-LGAD technology, allowing ePIC forward region to have a substantial overlap in the PID phase-space. In the barrel region similar AG-LGD based TOF and high performance DIRC (hpDIRC) technologies have been chosen for particle identification; also these two technologies have substantial phase-space overlap in the barrel region. In the electron going direction, a proximity focusing RICH (pfRICH) is chosen as the baseline detector. The Cherenkov photons for pfRICH are identified with High Rate Picosecond Photon Detector (HRPPD) produced by INCOM. Using the timing information of the Cherenkov photons produced at the window of the detector by the through going charged particles, the HRPPD sensors are capable of providing timing information. The pfRICH can provide pion-kaon separation up to 7~GeV/c. A schematic of the ePIC detector and PID devices are shown in figure \ref{fig:epic_scheme}.

\section{Proximity focusing RICH}      
The ePIC proximity focusing RICH uses a large proximity gap of about 40~cm. A layer of aerogel as a Cherenkov radiator is used to produce Cherenkov photons and they are detected by HRPPD photosensors produced by INCOM$^{TM}$ \cite{incom}. The HRPPDs are Micro Channel Plate-Photomultiplier tubes (MCP-PMTS) coupled to a 5~mm window coated with multialkali (K$_2$NaSb) photocathode that is used in transmission mode. The HRPPDs provide an active area of about 104 mm. HRPPDs are essentially derivatives of previous generation LAPPD (\cite{lappd}) type detectors produced by INCOM. The HRPPD readout is segmented in 32 times 32 pads of thickness 3~mm and pitch 3.25~mm. Detailed simultion studies have been performed to estimate Cherenkov angle resolution to be about 5~mrad. In the baseline design one layer of aerogel has been chosen with nominal refractive index 1.04, providing on an average 12 detected photons allowing to achieve pion kaon separation above 7~GeV. The design of the pfRICH consists of conical mirrors placed along the vessel wall and the beam pipe to detect photons which are otherwise lost. These mirrors therefore allow to increase the acceptance of the pfRICH.

A detailed simulation study has been made for microscopic tuning of the pfRICH, allowing an optimization of the parameters for the aerogel, the photo-sensors, the mirrors etc. The simulation studies are ongoing for risk mitigation, detector performance evaluation and setting benchmark plots for the upcoming beam tests with the prototype design of the pfRICH.

The HRPPDs are intrinsically fast photon detectors, with $\sim$ 50~ps single photo-electron time resolution. On av average about 80 Cherenkov photons that are produced in the window of the HRPPD photon-sensors are detected, and they provide excellent time information (~20 ps) for the through-going charged particles. This allows for additional time separation of pion and kaons flight and serves as a timing detector.  
     
\begin{figure}[!thb]
	\includegraphics[width=0.95\textwidth]{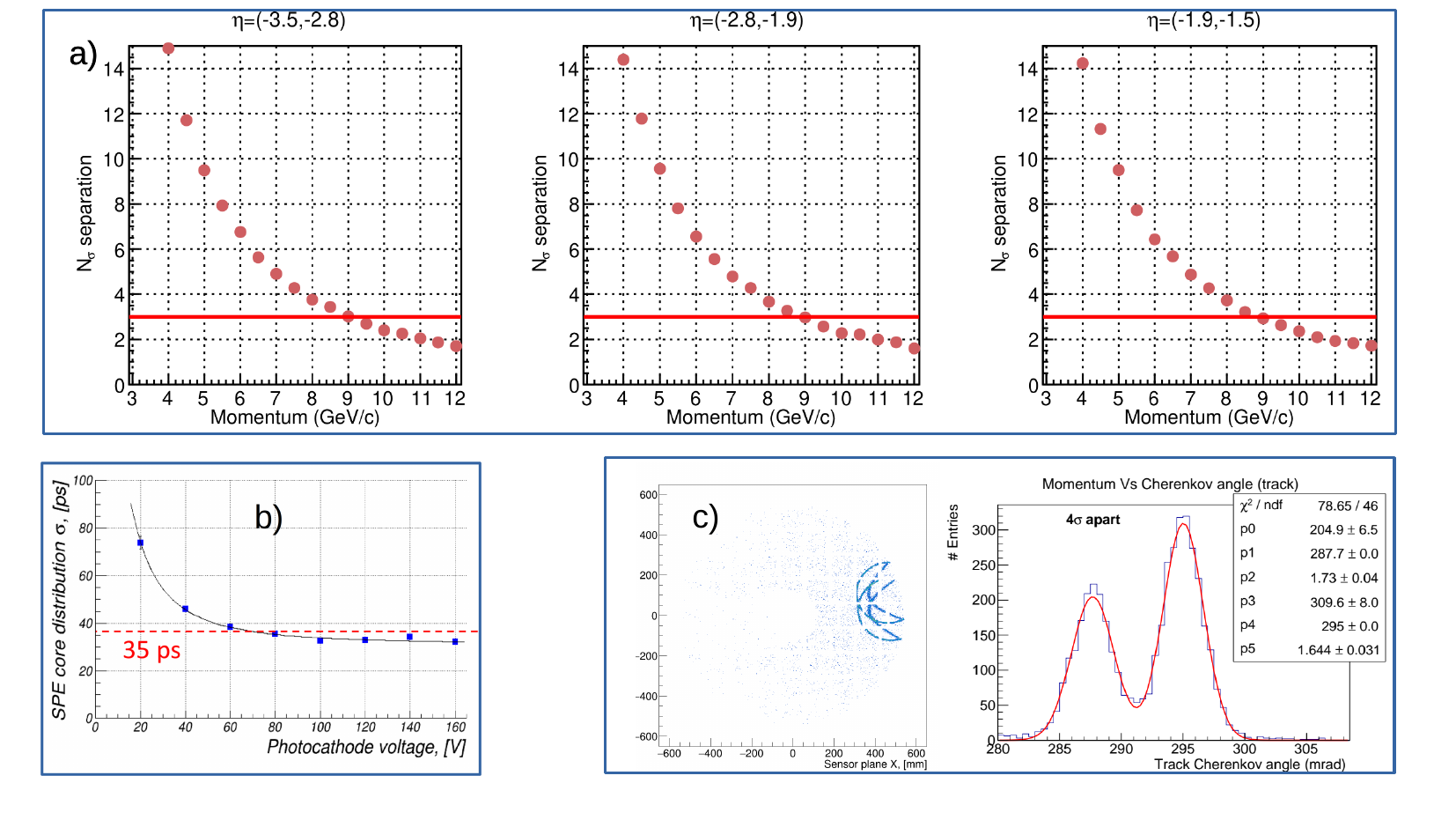}
	\caption{Some performance features of pfRICH; a) Number of sigma separation for pions and kaons as a function of momentum. With pfRICH design, one can achieve the EIC YR requirements, b) Time distribution of HRPPD sensors for single photo electron detection (SPE). Better than 50~ps time resolution for SPE distribution is seen, c) Sophisticated chi-square algorithm allows to provide excellent separation for complicated event topologies. }
	\label{fig:pfRICH}
\end{figure}

Extensive hardware activities are ongoing for the pfRICH; these studies include, a detailed characterization and optimization of the aerogel parameters, optimization of the mirror coating systems using thermal evaporator methods, design and optimization of pfRICH vessel  and most importantly with the photo-sensors. The HRPPD sensors will be placed in the spectrometer where the solenoidal field can be as high as 1.5T. Therefore, an important aspect has been to characterize the performance of such photon detectors in presence of an external magnetic field. Several tests have been performed at Argonne National Laboratory, USA and CERN with the LAPPD sensors to characterize their performance in presence of a magnetic field. Dedicated beam test has been performed to validate the timing performance of the LAPPD detectors. Being a fast photon detector, the HRPPD also aims to provide timing information for the thoroughgoing charged particle by detecting Cherenkov photons generated at the window of the photo-sensors. From the test beam studies at CERN it is expected to have about 80 ps timing resolution for single Cherenkov photons. A summary of the pfRICH performance is shown in figure \ref{fig:pfRICH}.
\section{A dual radiator RICH }
A dRICH aims to employ two radiators to cover large momentum region with substantial overlap. The baseline dRICH has been designed with aerogel with nominal refractive index of 1.02, and C$_2$F$_6$ gas as a radiator with nominal refractive index 1.008. SiPM based photo-sensors are used and distributed in six sectors to detect focalised photons reflected by their own six $60^\circ$ spherical mirror sectors. In the baseline design a 120~cm vessel length along the beam line is considered. The aerogel thickness is 4~cm. The dRICH aerogel starts at 198~cm from the interaction point (IP), with 90~cm radial size. This allows one to go down to $\eta\sim$1.5 in acceptance. The outer radius of the dRICH is 180~cm. This allows to have almost no absorption of photons in the vessel boundary for the photons coming from high pseudo-rapidity particles. The  An extruded sensor box of radius 185~cm is considered to host the spherical sensor planes with 10 cm attached services (including Front-end electronics, cooling and back-end services) covering the entire acceptance of dRICH. Each Photon Detection Unit (PDU) is grouped into 2X2 matrix of photosensitive surfaces with a 0.2~mm gap. Each sensor surface contains 4X4 SiPM pixels of 3~mm pitchs \cite{sipm}. Therefore, a total of 64 pixels are present in a PDU. Realistic parameters for the Rayleigh scattering, absorption length of the aerogel has been obtained from CLAS12 aerogel parameters for the nominal design (parameters have been taken from \cite{clas12}). For the simulation studies the parameterization of the refractive index of  C$_2$F$_6$ is taken from \cite{C2F6ref1}. 

\begin{figure}[!thb]
	\includegraphics[width=0.95\textwidth]{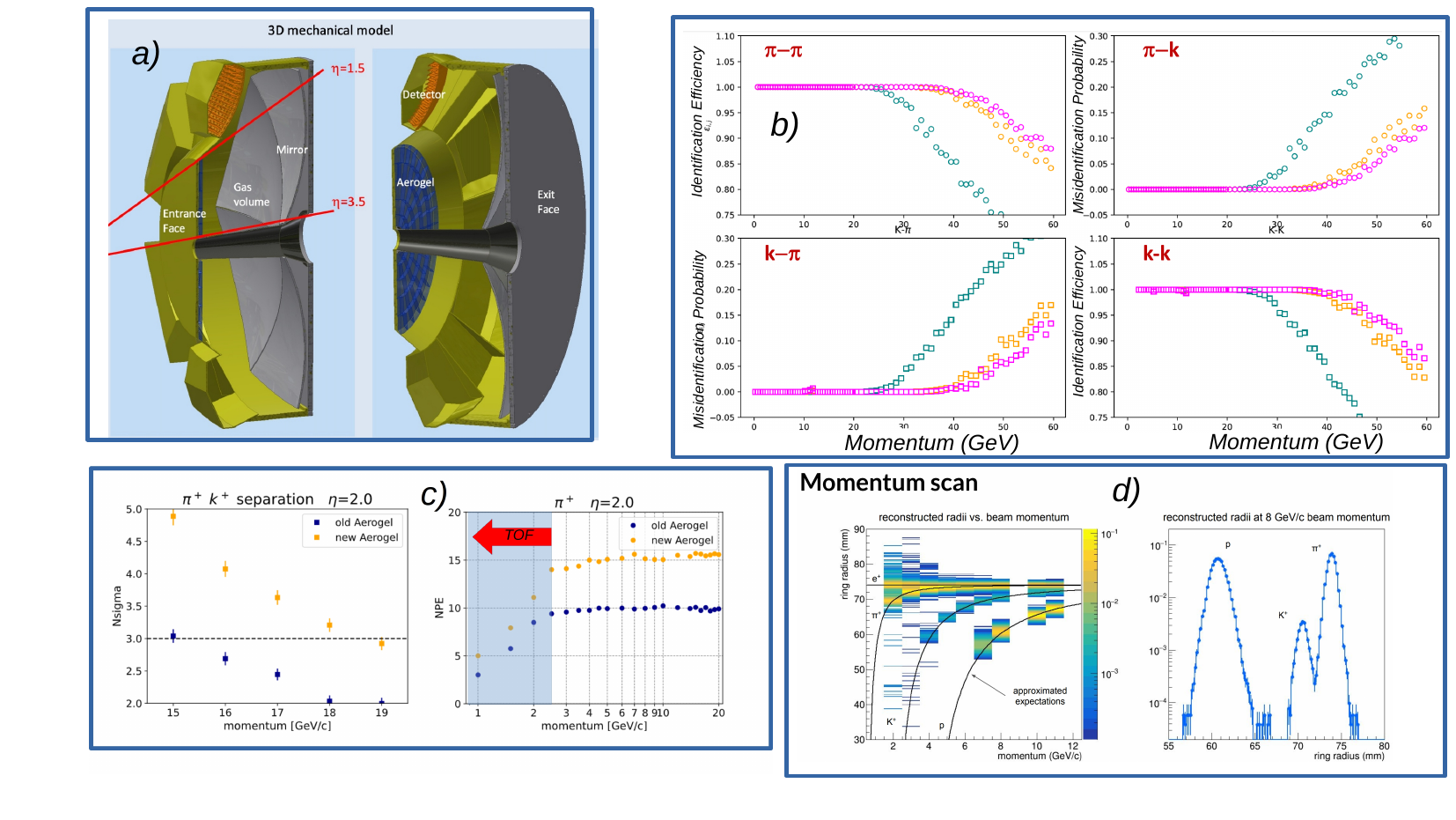}
	\caption{dRICH in ePIC: a) The mechanical designs demonstrate the aerogel and gas radiator, the sensor planes and the mirrors. In the official simulation framework, almost identical design has been implemented, b) the dRICH PID performance using information of Cherenkov photons emitted in the aerogel and gas. In the most critical pseudorapidity region (between 2.0 to 3.5) the dRICH has 95\% pion identification efficiency; c) different aerogel parameters are under study, aerogel with better optical qualities (referred as new) has substantially larger number of detected photons and performance; d) ongoing data analysis with the dRICH prototype in a test beam demonstrates pion kaon and proton rings are well separated as a function of momentum.}
	\label{fig:dRICH}
\end{figure}

The optimization of the optical parameters for the aerogel is also ongoing to push the PID limit at the lower momentum. Aerogel samples with higher refractive index and better optical parameters are under investigation; to increase the number of detected photons and also to improve the single photon resolution. Different samples of aerogel has been characterized in the laboratory and the parameters have been used to study the separation of pions and kaons samples in the dedicated simulation studies. Extensive laboratory studies have been performed with the SiPM photo-sensors to mitigate the dark count rates, also it has been demonstrated that cyclic annealing methods at high temperature has allowed the SiPMs to sustain their radiation hardness \cite{sipmRP}. Test beams exercises have been performed to validate the dRICH performance. The results are consistent with the simulation studies. In figure \ref{fig:dRICH} is it shown that the optimized configuration of dRICH is capable to achieve a comfortable momentum overlap with the aerogel, and a 3$\sigma$ $\pi$/K separation up to $\sim$50~GeV/c in most demanding high pseudo-rapidity region.  
\section{A high performance DIRC }
A detector based on the Detection of Internally Reflected Cherenkov light (DIRC) principle, with a compact radial size of only 7-8 cm, was selected as the PID system for the ePIC barrel region. It is anticipated to achieve a $\pi$/K particle separation with a minimum significance of 3 standard deviations for momenta up to 6 GeV/c. This performance represents an improvement of nearly twice the current state-of-the-art DIRC systems, primarily attributed to the development of new 3-layer focusing lenses, small pixel sensors, and high-speed readout electronics.

\begin{figure}[!thb]
	\includegraphics[width=0.95\textwidth]{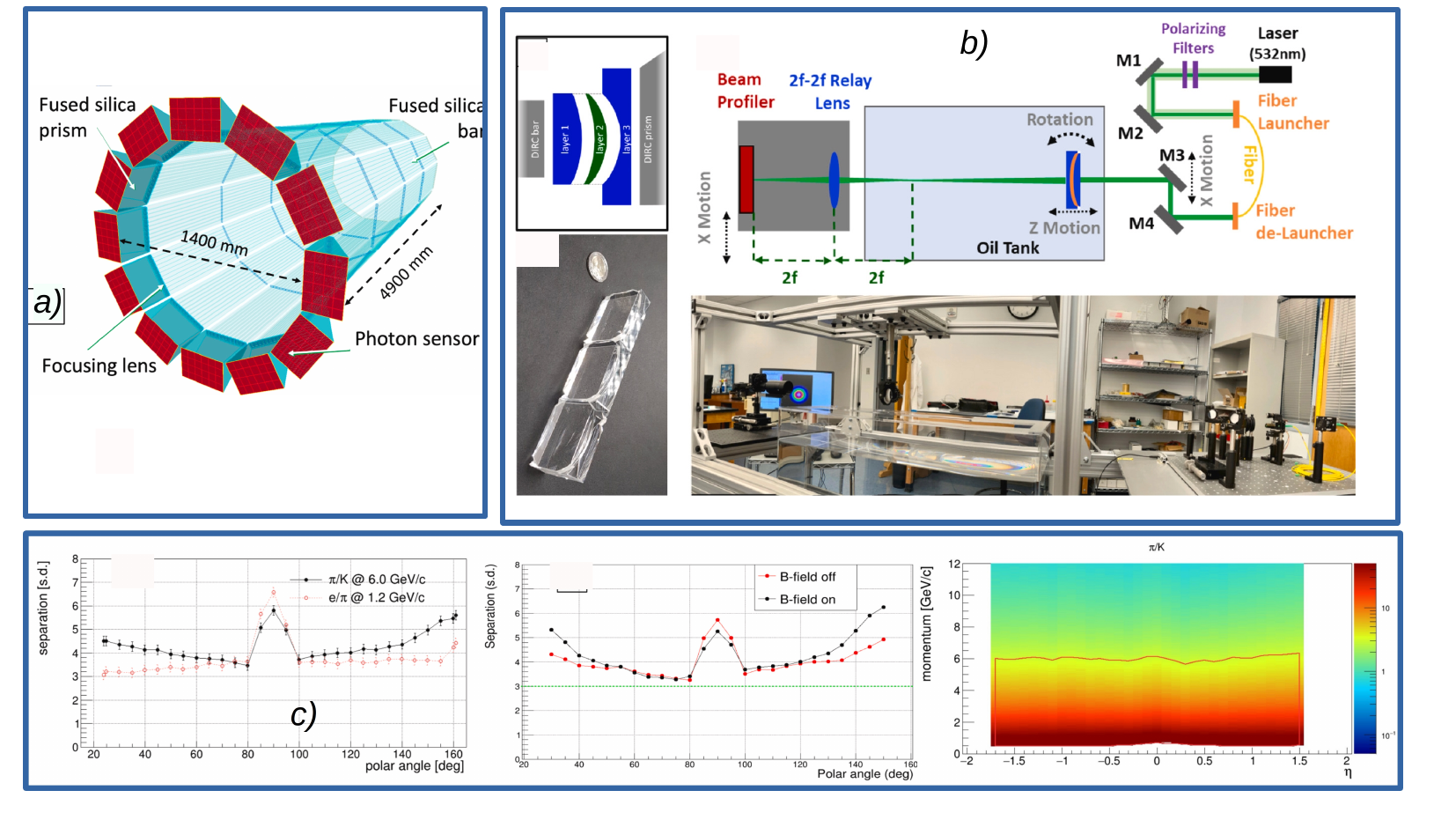}
	\caption{Panel a) shows the 3D model of the hpDIRC, the individual components are shown with hpDIRC dimensions;b) the scheme of the lens system and the laser setup to map out the focal plane of the 3-layer lens as a function of rotation angle of the lens is shown; c) Expected separation power performance of the hpDIRC as a function of the particle polar angle for electron-pion and pion-kaon hypothesis (left plot); in the middle plot, the effect of the magnetic field in the separation power is shown and the right most plot demonstrates the 3 standard deviation pion-kaon separation contour (in Z axis) in momentum pseudo-rapidity plane}
	\label{fig:hpDIRC}
\end{figure}

The high-performance DIRC (hpDIRC) is built upon the concept of a focusing DIRC detector, incorporating precise 3D (x, y, t) reconstruction. It is partitioned into twelve optically isolated sectors, each consisting of two light-tight containers, known as a bar box and readout box, which encircle the beamline within a 12-sided polygonal barrel with a slightly greater than 70 cm inner radius. Each bar box is equipped with a series of ten radiator bars, separated by small air gaps, made of synthetic fused silica. These bars are 4600 mm in length and have a cross-sectional dimension of 17 mm × 35 mm. At one end of each bar, mirrors are affixed to redirect Cherenkov photons toward the readout end, where they exit the bar and are focused, by a 3-layer spherical lens, on the back surface of a prism that serves as an expansion volume. The prism is constructed from synthetic fused silica and possesses an opening angle of 32 degrees, with dimensions of 237 mm × 350 mm × 300 mm. Each prism's backplane is covered with sensors featuring a pixel size of approximately 3 mm × 3 mm, enabling the recording of Cherenkov photons' position and arrival time. The designated sensor solution for the hpDIRC's baseline design consists of commercial microchannel plate PMTs (MCP-PMTs) with a pore size of 10 µm or less, available from Photek or PHOTONIS.
A comprehensive standalone Geant4 simulation was developed together with the PANDA Barrel DIRC group and validated with particle test beam data. The simulation incorporates realistic geometry and material properties. It accounts for background effects stemming from hadronic interactions and delta electrons within the radiator bars, as well as contributions from MCP-PMT dark noise and charge sharing among anode pads. All significant sources of resolution affecting hpDIRC performance were included, with Gaussian smearing applied to represent conservative resolution assumptions. It was assumed that the sensor and readout electronics collectively provide a timing precision of 100 ps per photon, and the ePIC tracking system's resolution at the hpDIRC radius was set at 0.5 mrad. Figure \ref{fig:hpDIRC} demonstrates the summary of simulation performances, ongoing lab activities etc. Detailed simulation studies concerning the DIRC's performance and several experimental measurements of key components have been conducted and documented in reference \cite{hpdirc}. 

\section{Time of Flight}
\begin{figure}[!thb]
	\includegraphics[width=0.95\textwidth]{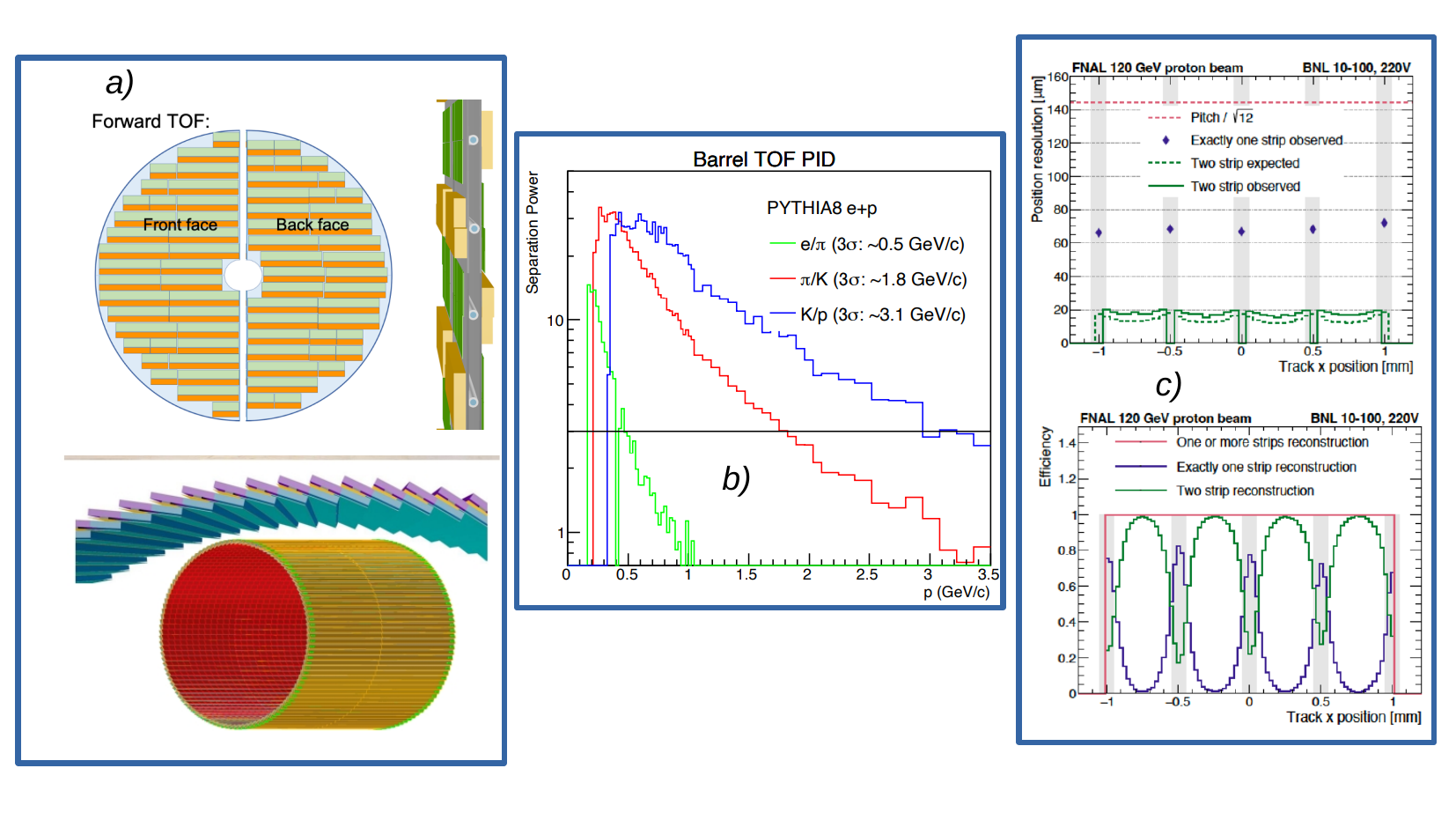}
	\caption{We see the model of the Forward and the barrel TOF in panel a); an example of the separation power of the barrel tof is shown in panel b); in panel c) we see the position resolution and the geometric efficiency of AC-LGAD from FNAL test beam data. }
	\label{fig:tof}
\end{figure}

An AC coupled Low Gain Avalanche Diode (ACLGAD) serves both the purpose of tracking and performing Time Of Flight (TOF) PID in both barrel and the forward region. The fabrication of the AC-LGAD requires a thin layer of highly resistive silicon p-type semiconductor substrate, a large n+ shallow implant is then used to cover the deep p+layer for the p-n junction. On top of this, the electrodes of the sensors to connect to the readout electronics are dielectric metal pads which are separated from the n+layer by a thin insulator. This coupling allows the adjacent readout channels to share the charge, hence improving the pixel hit resolution. The barrel region the AC-LGAD has to cover about $\sim$11m$^{2}$ of area, whereas in the forward region the coverage is about 2.2m$^{2}$. The TOF detector will have substantial overlap with the hpDIRC and the dRICH. A strip readout (500~$\mu$m x 1~mm) has been considered. It has been demonstrated that 100\% fill-factor can be achieved if two or more strips are used for the reconstruction \cite{aclgad}. Pions and kaons can be separated at 3$\sigma$ limit u to 2~GeV and electrons and pions can be separated up to 500~MeV in the barrel region. In the forward region also AC-LGAD based time of flight detector is considered, figure \ref{fig:tof} summarizes the performance of tof in barrel region as suggested by extensive simulation studies and performance of ACLGAD from the data analysis of the beam test. Contrary to the barrel region, a (500~$\mu$m x 500~$\mu$m) pixel read-out is considered. In the forward region 3-$\sigma$ pion-kaon and electron-pion separation can be achieved up to 2.7~GeV and 800~MeV respectively. 

\end{document}